\documentclass[pra,twocolumn,superscriptaddress,showpacs,amsmath,amstex,amssymb,citeautoscript]{revtex4-1}

\usepackage[T1]{fontenc}
\usepackage[utf8]{inputenc}
\DeclareUnicodeCharacter{2212}{\textminus}
\usepackage{lipsum}
\usepackage{amsmath}
\usepackage{amssymb}
\usepackage{bbm}
\usepackage{braket}
\usepackage{xcolor}
\usepackage{pifont}
\usepackage[mathscr]{euscript}
\usepackage[shortlabels]{enumitem}
\usepackage{graphicx}
\usepackage{dsfont}
\usepackage{comment}
\usepackage[colorlinks=true]{hyperref}  
\hypersetup{
    bookmarks=true,         
    unicode=false,          
    pdftoolbar=true,        
    pdfmenubar=true,        
    pdffitwindow=false,     
    pdfstartview={FitH},    
    pdftitle={Probing pairing symmetries through quasiparticle interference in chiral Bloch bands},    
    pdfauthor={Banerjee, Mandal, Orth, Scheurer},     
    pdfsubject={},   
    pdfcreator={},   
    pdfproducer={}, 
    pdfkeywords={} {} {}, 
    pdfnewwindow=true,      
    colorlinks=true,       
    linkcolor=blue, 
    citecolor=blue,        
    filecolor=magenta,      
    urlcolor=blue           
}

\newcommand{\equref}[1]{Eq.~(\ref{#1})}

\newcommand{\secref}[1]{Sec.~\ref{#1}}
\newcommand{\figref}[1]{Fig.~\ref{#1}}
\newcommand{\refcite}[1]{Ref.~\onlinecite{#1}}

\newcommand{\tableref}[1]{Table~\ref{#1}}
\newcommand{\appref}[1]{Appendix~\ref{#1}}

\newcommand{\pdagger}{{\phantom{\dagger}}}

\renewcommand{\vec}[1]{\boldsymbol{#1}}

\definecolor{wrongultramarine}{rgb}{1,0.5,0}
\begin{document}

\title{Probing pairing symmetries through quasiparticle interference in chiral Bloch bands}

\author{Sayan Banerjee}
\thanks{These two authors contributed equally.}
\affiliation{Institute for Theoretical Physics III, University of Stuttgart, 70550 Stuttgart, Germany}

\author{Subrata Mandal}
\thanks{These two authors contributed equally.}
\affiliation{Institute for Theoretical Physics III, University of Stuttgart, 70550 Stuttgart, Germany}

\author{Peter P.~Orth}
\affiliation{Department of Physics, Saarland University, 66123 Saarbrücken, Germany}
\affiliation{Center for Quantum Technologies (QuTe), Saarland University, 66123 Saarbrücken, Germany}

\author{Mathias S.~Scheurer}
\affiliation{Institute for Theoretical Physics III, University of Stuttgart, 70550 Stuttgart, Germany}

\begin{abstract}
Recent experiments in van der Waals multi-layer systems have demonstrated that superconductivity can emerge from symmetry-reduced, chiral normal states. We here provide a theory for quasiparticle interference (QPI) of superconductors with chiral Bloch bands. 
Our analysis reveals how the non-trivial quantum geometry of the Bloch states crucially affects the interference pattern even in the normal state, inducing significant sublattice dependence. In the superconducting state, the behavior becomes more complex due to the interplay of the quantum geometry of the Bogoliubov quasiparticles with the momentum-dependent phase of the order parameter. 
We reveal how the spatial dependence of the local spectral function around impurities can be used to distinguish between different candidate pairing states, both with zero and finite center-of-mass momentum. Our work thus provides guidance to interpreting QPI patterns in materials with chiral bands, which  may be useful when probing 
the rich physics of pairing in such systems.
\end{abstract}

\maketitle

\section{Introduction}
Scanning tunneling microscopy (STM) is a powerful and broadly applied experimental technique to probe materials~\cite{Binnig1983Mar,RevModPhys.59.615,Yin2021Apr} since it provides access to local electronic properties, such as the local electronic spectral function. However, the locality of the tunneling process effectively leads to an average over crystalline momenta, precluding direct access to momentum-resolved information. Accordingly, in a superconductor, for example, one has access to the density of states of the Bogoliubov quasiparticles but cannot directly extract the momentum dependence of the gap or of the phase of the superconducting order parameter. The complex nature of the order parameter is one of the main reasons why identifying the pairing symmetry of a superconducting system is generally a challenging task: order parameters with different symmetries might have the same spectral function, such as the $s$-wave state $\Delta_{\vec{k}} = \text{const.}$ and the chiral $p$-wave order parameter $\Delta_{\vec{k}}\propto k_x + i k_y$ (for an isotropic Fermi surface), since the spectrum is only sensitive to $|\Delta_{\vec{k}}|$.

One way to use STM to gain access to momentum-resolved quantities such as the dispersion and form factors of the electronic states or the phase of the superconducting order parameter is by probing quasiparticle interference (QPI)~\cite{Crommie1993Jun, byersInfluenceGapExtrema1993, Hoffman2002Aug, mcelroyRelatingAtomicscaleElectronic2003, QPIgeneral2}. QPI has become a leading experimental method to probe properties of quasiparticles in quantum materials and has been applied to a variety of material classes such as conventional and unconventional superconductors~\cite{Hoffman2002Aug, Wang2003QPI,PhysRevLett.108.127001,Balatsky2006RMP,allanAnisotropicEnergyGaps2012,allanImagingCooperPairing2013,sprauDiscoveryOrbitalselectiveCooper2017, nagSCsymmetry,Wang_2025, jiang2026nonreciprocalimpurityscatteringprobe,PhysRevB.105.064504,PhysRevB.92.184513,akbari2013full,jmmf-mpr8,pttl-8m71}, including inelastic tunneling \cite{chi2025quasiparticleinterferencelifeassignature}, correlated charge-density wave materials ~\cite{arguelloQuasiparticleInterferenceQuasiparticle2015, kunduLowenergyElectronicStructure2024}, and alter- and $p$-wave magnets \cite{zhangCrystalsymmetrypairedSpinValley2025, AMQPI,AMQPI2,AMQPI3,Petermann_2025}. QPI has also been used to elucidate the nontrivial topology and quantum geometry of Bloch states, for example, of Dirac fermions~\cite{PhysRevLett.101.206802, roushanTopologicalSurfaceStates2009,BerryphaseQPI, GrapheneQPI, nguyen2025glimpsingelectronsformfactor, Chen_2017,PhysRevB.98.075115} and Weyl semimetals~\cite{PhysRevB.98.075115,batabyalVisualizingWeaklyBound2016, inoueQuasiparticleInterferenceFermi2016, WeylSemimetals, WeylSemimetals2}.

QPI is based on studying the spatial \emph{variation} of the local spectral function in the vicinity of impurities. The presence of inhomogeneities leads to scattering between momentum eigenstates of the homogeneous limit of the underlying quantum system; the Fourier transform of the resulting spatial dependence of the interference effects is crucially determined by the allowed scattering wave vectors and, thus, by the momentum-dependence of the quasiparticle spectra. 
Importantly, since the measured QPI signal depends in general in a complicated way on a combination of kinetic and wavefunction effects associated with the Bloch states, a detailed theoretical analysis is needed to reliably extract information from experimental data. In addition, 
interpreting QPI patterns often proves challenging due to the dependence on microscopic details, an issue that has already been noticed and addressed theoretically in early works of the field~\cite{PhysRevB.68.014508,PhysRevB.68.180506,PhysRevB.67.100506,PhysRevB.67.172505,PhysRevB.73.104511,PhysRevB.86.134509,PhysRevB.69.060503,PhysRevLett.114.217002}. 

These wavefunction or ``quantum geometry’’~\cite{ReviewQGBernevig} effects are particularly prominent in multi-band systems with low symmetries. One such system of current interest is rhombohedral tetralayer graphene (RTG), which lacks any form of intravalley anti-unitary symmetry or reflection symmetry, such that the Bloch states of a single valley are chiral and exhibit a finite net Berry curvature \cite{koshino2009,koshino2010interlayer,PhysRevB.82.035409,min2008pseudospin,koshino2010parity,muten2021exchange,AliceaQuantumGeometry,jahin2025enhancedkohnluttingertopologicalsuperconductivity,BerryTrashcan,dealmeida2026highharmonicgenerationsystemschiral}. In addition, there are significant interaction effects \cite{SCTrilayer,HalfQuarterMetals,NontwistedReview,AcousticPhononsDasSarma,BergRGKohnLuttinger,YiZhuangKohnLuttinger,BitanRoy,ShubhayusPaper,MultilayerGeneralSerbyn,PacoBiAndTri,MacDonaldFRG,Paco2023,Chubukov,DongSpinCanting}, which can also lead to the spontaneous polarization of the two valleys that can, in turn, coexist with superconductivity \cite{rtg_chiral_sc_experiment,JiaSixLayers}. As discussed in a variety of theoretical works \cite{PhysRevB.111.174523,2026NatCo..17..232G,StandfordWithTheRelativeChirality,Qin2026Mar,Paco,fcdc-9lm3,YahuisPaper,fdz1-dbf6,sedov2025probingsuperconductivitytunnelingspectroscopy,OurEnergetics,2025arXiv251019943S,zgnk-rw1p,FranzSCRingOfFire,k8sb-rqxf,PatrickLeeControl,2026PhRvB.113g5154G,2025arXiv250608087L,pqq5-3n84,2025arXiv251020816A,2025arXiv250916312L,2q6v-4brs}, this has the potential to lead to fascinating superconducting properties. There are still multiple possible candidate pairing states and developing techniques to (uniquely) identify them (or at least narrow down potential candidates) and to probe the microscopic coexistence of superconductivity with valley imbalance are urgently needed. Similar behavior of pairing that emerges from a valley-imbalanced normal state was also reported for twisted MoTe$_2$ \cite{MoTe2_unconv_exp}.

Motivated by the rich phenomenology of these systems, we here study QPI in the valley-polarized normal state and in different possible superconducting phases for a prototype model of chiral valleys. The model we study applies directly to RTG but is also expected to capture the main features of twisted MoTe$_2$. Our focus will be on scrutinizing the non-trivial effects coming from the quantum geometry of the normal-state Bloch states and their interference with the momentum dependence of possible superconducting order parameters. We also investigate under which conditions QPI provides a way to pinpoint the form of the superconducting order parameter. We note that the complementary strong-tunneling regime, which also provides phase-sensitive information about pairing via Andreev-tunneling processes, was studied in \refcite{sedov2025probingsuperconductivitytunnelingspectroscopy} for valley-polarized RTG.

The remainder of the paper is organized as follows. In \secref{sec:ModelDefinition}, we define the model we use. The resulting QPI patterns for the chiral normal state are discussed in \secref{QPINormalState}. Section~\ref{QPIInSuperconductor} deals with the behavior for different superconducting candidate orders. The results are summarized in \secref{ConclusionSec} and three appendices provide further numerical results and details on the calculations.

\section{Model for Normal state}\label{sec:ModelDefinition}
To study QPI for chiral Bloch states, we take the example of rhombohedral tetralayer graphene (RTG). This corresponds to ABCA stacking where the four graphene monolayers are each displaced relative to the previous one by the nearest-neighbor lattice vector. Throughout this work, we focus on the valley-polarized regime of RTG~\cite{rtg_chiral_sc_experiment}, i.e., we only need to take into account the bands of one of the two valleys when investigating the low-energy behavior around the Fermi level. We note that a similar analysis 
would apply for twisted MoTe$_{2}$~\cite{MoTe2_unconv_exp}, but here we elucidate the key findings using the RTG model. 

While the full tight-binding Hamiltonian for RTG corresponds to an 8-band model per spin (four layers combined with two sublattices), we here restrict ourselves to an isotropic two-band minimal model in the continuum limit, retaining only the $A_1$ and $B_4$ sublattices which are the dominant low-energy 
degrees of freedom \cite{koshino2009,koshino2010interlayer,PhysRevB.82.035409,min2008pseudospin,koshino2010parity,muten2021exchange}. Despite its simplicity, this model captures the essential low-energy physics of RTG, including the band flattening that is likely relevant for the emergence of correlation effects including superconductivity. 
In the following, we will refer to the $A_1$ and $B_4$ sublattices as the top (t) and bottom (b) layers, respectively.

Introducing electron annihilation and creation operators $c_{\vec{k}\alpha}$ and 
$c^{\dagger}_{\vec{k}\alpha}$ for momentum $\vec{k}$ and sublattice index 
$\alpha \in \{\text{t}, \text{b}\}$, the normal-state Hamiltonian reads
\begin{equation}
\begin{split}
H_0 &= \sum_{\vec{k}} c^{\dagger}_{\vec{k}\alpha} [h(\vec{k})]_{\alpha\beta}^{\pdagger} c_{\vec{k}\beta}^{\pdagger}, \\
h(\vec{k}) &= \begin{pmatrix}
 u_{0}-\mu& w_{0} (k_{x} + i k_{y})^4 \\
w_{0} (k_{x} - i k_{y})^4 & -u_{0}-\mu 
\end{pmatrix}.
\end{split}
\label{eq:NormalStateHamiltonian}
\end{equation}
Here $\mu$ is the chemical potential, which controls the carrier density. The 
parameter $u_0$ encodes the layer-asymmetry energy induced by a perpendicular 
displacement field, while $w_0$ is the interlayer hopping amplitude. The system is invariant under $120^\circ$ rotation along the $z$ direction, denoted by $C_{3z}$ in the following. We choose the rotation axis of $C_{3z}$ through the $A_1$ ($t$) sublattice, as this is convenient to discuss pairing \cite{sedov2025probingsuperconductivitytunnelingspectroscopy}; $C_{3z}$ then acts as $c_{\vec{k},\alpha} \rightarrow e^{i\phi_\alpha} c_{C_{3z}^{-1}\vec{k},\alpha}$ on the momentum-space fermionic operators where $\phi_t=0$ and $\phi_b =  2\pi/3$.

We compute the QPI response using the eigenbasis of $h(\vec{k})$. To this end, we choose the phases of the Bloch eigenstates $|v^{p}_{\vec{k}} \rangle$ such that 
\begin{equation}
  |v^{p}_{\vec{k}} \rangle = N_{\vec{k}} \begin{pmatrix}
  u_0 +p \sqrt{u_{0}^2 + w_{0}^2\vec{k}^8}  \\
  w_0 (k_x - i k_y)^4
  \end{pmatrix}
  \equiv 
  \begin{pmatrix}
A_{\vec{k}} \\ B_{\vec{k}}
\end{pmatrix},
  \label{eq:BlochEigenstates}
\end{equation}
where $N_{\vec{k}} \in \mathbb{R}^{+} $ is a normalization factor and $p = \pm$ labels the 
two low-energy bands; in this gauge, the Bloch states of the bands transform trivially under 
the aforementioned $C_{3z}$ symmetry. The resulting dispersion of the two low-energy bands is given by 
$E^p_{\vec{k}} = p\sqrt{u_{0}^2 + w_{0}^2\vec{k}^8} - \mu$ and is
shown in \figref{fig:NormalStateQPI}(i) for a representative choice of parameters.

\section{QPI in normal state}\label{QPINormalState}
In this section, we analyze the impurity-induced spatial modulation of the local density of states (LDOS) (i.e.~the Friedel oscillations) in the
normal state using the continuum model introduced in \secref{sec:ModelDefinition}. We scrutinize the
effects of quantum geometry, entering through the momentum dependence of the Bloch states $|v^p_{\vec{k}} \rangle$.
The Fourier transform of the real-space LDOS modulation defines the QPI signal measured in Fourier-transform scanning tunneling spectroscopy (FT-STS) \cite{Balatsky2006RMP,Wang2003QPI}.
Since we here focus on real-space signatures, we present the corresponding LDOS modulations directly in
real space, and later discuss their interpretation in terms of QPI patterns.

We model isolated (nonmagnetic) point-like impurities as local scattering potentials
$V_{\vec{k}\alpha,\vec{k}'\beta}$ that scatter an electron from momentum $\vec{k}'$ and layer index $\beta$ to
momentum $\vec{k}$ and layer index $\alpha$. The corresponding impurity Hamiltonian reads
\begin{equation}
H_{\text{imp}}=\dfrac{1}{N} \sum_{\vec{k},\vec{k}'} c^{\dagger}_{\vec{k} \alpha} V_{\vec{k} \alpha, \vec{k}' \beta} c_{\vec{k}' \beta}^{\pdagger},
\end{equation}
where $N$ is the number of Bravais lattice sites. 
We assume that the impurity is localized at $\vec{r}=0$ on either the top or bottom layer, such that $V_{\vec{k} \alpha, \vec{k}' \beta} \neq 0$ only if $\alpha = \beta = \text{t}$ for an impurity on the top layer or if $\alpha = \beta = \text{b}$ on the bottom layer. For concreteness, we will further focus on momentum independent $V_{\vec{k} \alpha, \vec{k}' \alpha}$ such that we are left with two inequivalent impurity types, described by
\begin{equation}
V_{\vec{k} \alpha, \vec{k}' \beta} = V_0
\begin{pmatrix}
1 & 0 \\
0 & 0
\end{pmatrix} \equiv M_{\text{t}} , \ 
  V_{\vec{k} \alpha, \vec{k}' \beta}= V_0
\begin{pmatrix}
0 & 0 \\
0 & 1
\end{pmatrix} \equiv M_{\text{b}}.
\label{eq:ImpurityMatrices}
\end{equation}
These encompass scattering processes acting exclusively on the $A_1$ (top) and $B_4$ (bottom) sublattices,
respectively. Due to the spatial separation of the first and fourth layers (there are two graphene layers in between), it makes sense to focus on impurities that only impact one of the two layers at a time. While breaking translational symmetry, both impurity potentials in \equref{eq:ImpurityMatrices} respect the aforementioned $C_{3z}$ symmetry.

To compute the impurity-induced change in the local response, we first introduce the retarded Green's
function of the clean system at energy $\omega$ and momentum $\vec{k}$,
\begin{equation}
G^R_0(\vec{k},\omega) = \sum_p \dfrac{|v^p_{\vec{k}} \rangle \langle v^p_{\vec{k}}|}{\omega+i\eta- {E}^p_{\vec{k}}}
\label{eq:BareGF}
\end{equation}
where $|v^p_{\vec{k}} \rangle$ and $ {E}^p_{\vec{k}}$ are the Bloch eigenstate and energy for the $p$th
band, respectively, and $\eta>0$ is a phenomenological broadening parameter (kept fixed in what follows).
Working to leading order in the impurity potential (Born approximation), the impurity-induced correction to
the Green’s function in momentum space is
\begin{equation}
\delta G^R_{\vec{k} \alpha,\vec{k}'\beta}(\omega) = [G^R_0(\vec{k},\omega) M G^R_0(\vec{k}',\omega)]_{\alpha\beta}.
\end{equation}
Finally, we project onto a specific layer $|\alpha\rangle$ corresponding to the position of the STM tip,
i.e., the layer on which the local density of states is measured. In the present two-band model, the
projection states are simply $|\text{t} \rangle = (1,0)^T$ and $|\text{b} \rangle = (0,1)^T$. The resulting changes in the projected, impurity-induced Green's function at position $\vec{r}$ can then be written as
\begin{equation}
\begin{split}
& \delta G_{\alpha\beta}^R( \vec{r}, \omega)= \int d^2\vec{k} \int d^2\vec{k}' e^{i (\vec{k}-\vec{k}')\cdot\vec{r}} \\
& \times \sum_{p,p'} \langle \alpha |v^p_{\vec{k}} \rangle \dfrac{1}{\omega+i\eta - {E}^p_{\vec{k}}}  \langle v^p_{\vec{k}}| M_{\beta} | v^{p'}_{\vec{k}'}  \rangle \dfrac{1}{\omega+i\eta - {E}^{p'}_{\vec{k}'}} \langle v^{p'}_{{\vec{k}}'} | \alpha \rangle, 
\end{split}
\label{eq:GFchangeNS}
\end{equation}
Here, the first index $\alpha\in\{\text{t},\text{b}\}$ specifies the measurement layer (STM tip), while the second index
$\beta\in\{\text{t},\text{b}\}$ specifies whether the impurity is on the top or bottom layer through the choice
$M_{\beta}\in\{M_\text{t},M_\text{b}\}$.

The corresponding change in the normal-state spectral function (and hence the LDOS modulation) is given by
\begin{equation}
    \delta A_{\alpha\beta}^{\text{NS}}(\vec{r}, \omega) = -\frac{1}{\pi} \text{Im}\left[\delta G^R_{\alpha\beta}(\vec{r}, \omega)\right].
    \label{eq:SpectralFunction}
\end{equation}
Since we consider an isotropic dispersion and impurity potential, it is sufficient to show one-dimensional cuts of
$\delta A_{\alpha\beta}^{\text{NS}}(\vec{r},\omega)$ obtained from \equref{eq:SpectralFunction} (choosing
$\vec{r}$ along the $x$-axis without loss of generality).

\begin{figure}[tb]
\centering
  \includegraphics[width=1.0\linewidth]{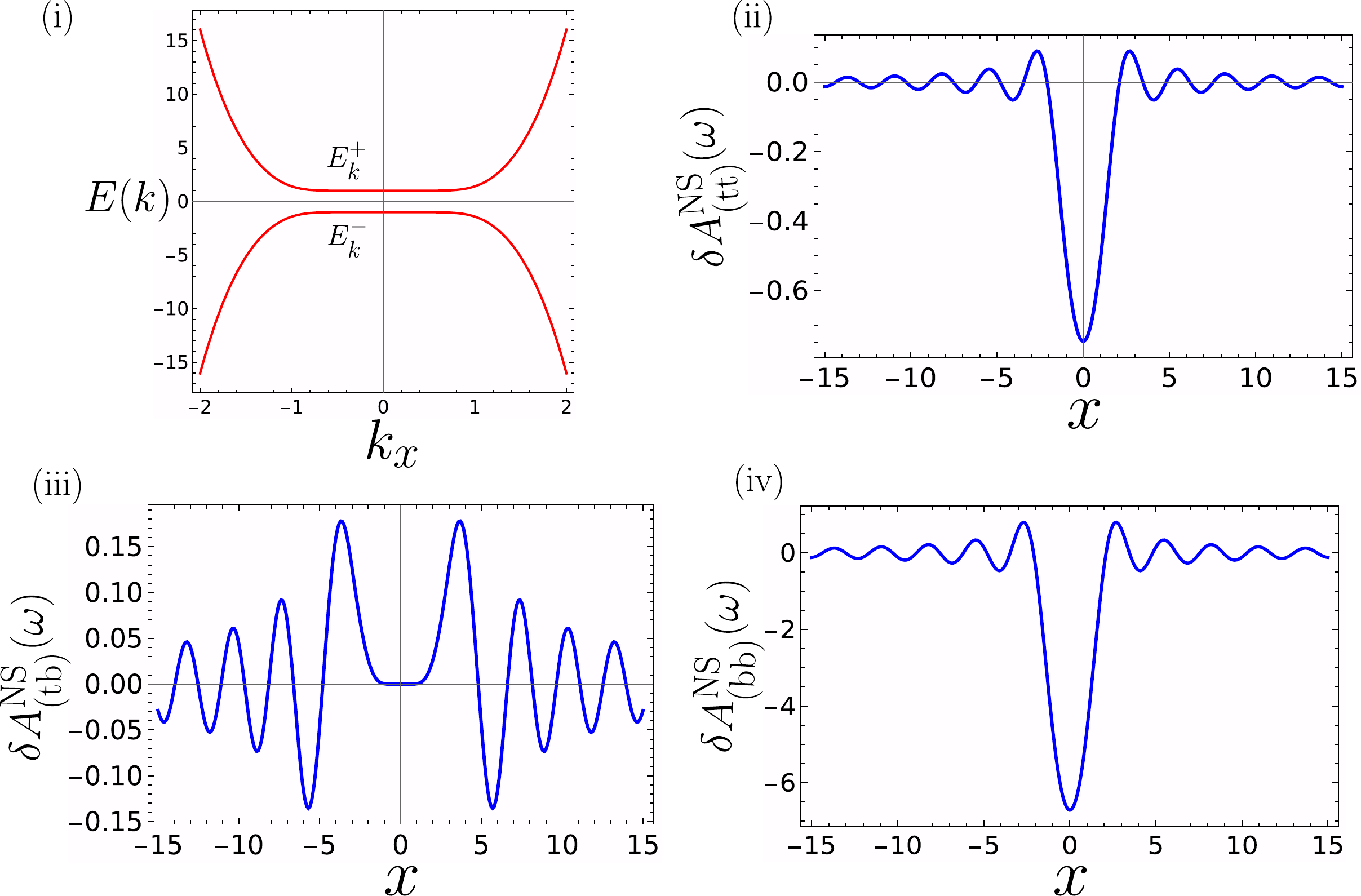}
  \caption{\textbf{Normal state results}. (i) Band structure of the low-energy Hamiltonian in \equref{eq:NormalStateHamiltonian} along the cut $\vec{k} = (k_x, 0)$, showing the upper ($E_{k}^+$) and lower ($E_{k}^-$) branches. One-dimensional cuts of the leading-order impurity-induced
  correction in the spectral function, $\delta A^{\text{NS}}_{\alpha\beta}(\omega)$ as a function of position $x$ for
  (ii) top-top (tt), (iii) top-bottom (tb), and (iv) bottom-bottom (bb) configurations. The bottom-top combination is identical to the top-bottom combination. We express momenta and lengths in units of $k^* = (u_0/w_0)^{1/4}$ and $1/k^*$, respectively, and energies in units of $u_{0}$. Here we set $u_{0}=w_{0}=1$, $\mu=0$, $\omega = -2$, $V_{0} =1$ and $\eta = 0.01$.}
  \label{fig:NormalStateQPI}
\end{figure}

In \figref{fig:NormalStateQPI}(ii), a sharp drop in $\delta A_{\text{tt}}^{\text{NS}}(\vec{r},\omega)$ at $x=0$ is observed for the configuration where both the STM tip and the impurity, which is placed at the origin, are located at the top layer ($\alpha \beta = \text{tt}$). This behavior arises because the impurity and the measurement point reside on the same layer, leading to a strong local response. Moving away from the impurity the usual spatially oscillatory Friedel modulations are observed, tapering away with distance from the impurity site, where the electronic states are less influenced by the impurity potential. 

On the other hand, the cross-layer combinations where impurity  and STM tip are on different layers exhibit a markedly different behavior as shown in \figref{fig:NormalStateQPI}(iii). In this case, the change in the spectral function $\delta A_{\text{tb/bt}}^{\text{NS}}(\vec{r}, \omega)$ vanishes at the origin, i.e. $\delta A_{\text{tb/bt}}^{\text{NS}}(\vec{0}, \omega) =0$. This surprising result is a direct consequence of the nontrivial quantum geometry of the Bloch states and can be readily derived from \equref{eq:GFchangeNS}, by noting that for $\vec{r} =0$, the change in the Green's function takes the form
\begin{equation}
    \delta G^{R}_{tb}(\vec{0},\omega) \propto \left|\int d^2 \vec{k}\sum_{p} \dfrac{ \langle t|v_{\vec{k}}^{p}\rangle \langle v_{\vec{k}}^{p}|b\rangle}{\omega+i\eta - {E}^p_{\vec{k}}} \right |^2.
\end{equation}
The relative momentum-dependent winding, $\propto e^{-4i\varphi_{\vec{k}}}$, of the lower and upper components of $\ket{v_{\vec{k}}^{p}}$ [cf.~\equref{eq:BlochEigenstates}] entering the integrand leads to $\delta G^{R}_{tb}(\vec{0}, \omega)=0$. This means that, as a result of quantum geometry-induced destructive interference, the impact of the impurity cancels out when the STM tip is directly above the impurity on the opposite layer.

For the configuration where both the impurity and the STM tip are on the bottom (bb) layer, the spatial dependence shown in \figref{fig:NormalStateQPI}(iv) reveals a  similar spatially oscillating qualitative behavior as $\delta A_{\text{tt}}^{\text{NS}}(\vec{r}, \omega)$. The primary difference between the tt and bb configurations is the magnitude of the response, related to the difference between $|A_{\vec{k}}|^2$ and $|B_{\vec{k}}|^2$ in \equref{eq:BlochEigenstates}. Taken together, these results indicate that the Bloch states strongly affect the QPI response, with the most significant impact in the off-diagonal tb/bt configuration displaying destructive interference when the STM tip probes the other layer right at the location of the impurity.

We have also checked that these conclusions remain qualitatively robust upon including trigonal warping,
which breaks the isotropy of the dispersion (see \appref{app:trigonalwarping}). In that case, the tt and bb configurations exhibit a six-fold
symmetric spatial pattern, while the cross-layer configurations show only three-fold rotational symmetry. The destructive interference in the mixed layer (tb and bt) configurations due to quantum geometry, which zeroes out the signal at the origin, also remains.

\section{QPI in superconducting state}\label{QPIInSuperconductor}
We now turn to the QPI response of the possible superconducting states \cite{MoTe2_unconv_exp,rtg_chiral_sc_experiment} that can emerge from the chiral normal state. Besides the momentum dependent phases between the two components of the Bloch states in \equref{eq:BlochEigenstates}, the momentum dependent phase of the superconducting order parameter will now also play a role.

\subsection{Basic setup}
To formulate the superconducting response, we project pairing and impurity scattering onto the active
low-energy band of the normal-state Hamiltonian $h(\vec{k})$ in \equref{eq:NormalStateHamiltonian}. Since scattering processes involving remote bands are suppressed by the inverse band gap,
we expect this projection to capture the essential low-energy physics of the superconducting state and its response to impurities. In our gauge choice, this band is described by
the Bloch eigenstate $|v^{+}_{\vec{k}}\rangle$ introduced in \equref{eq:BlochEigenstates}, with dispersion
$E_{\vec{k}}=\sqrt{u_0^2+w_0^2\vec{k}^8}-\mu$. We define the corresponding band operators as 
\begin{equation}
f_{\vec{k},\sigma}\equiv \sum_{\alpha=\text{t},\text{b}} \langle \alpha| v^{+}_{\vec{k}}\rangle c_{\vec{k}\alpha,\sigma},
\label{eq:bandProjection}
\end{equation}
where all subsequent pairing and scattering terms are implicitly projected onto this active band. For later convenience, we also define the two elements in the Bloch eigenstate $|v^+_{\vec{k}}\rangle 
\equiv (A_{\vec{k}}, B_{\vec{k}})^T$ where 
$A_{\vec{k}} \equiv N_{\vec{k}}\bigl(u_0 +
\sqrt{u_0^2+w_0^2 k^8}\bigr)$ and $  B_{\vec{k}} \equiv N_{\vec{k}} w_0 k^4 e^{-4i\varphi_{\vec{k}}}$ as in \equref{eq:BlochEigenstates}.

We consider both zero-momentum and finite-momentum  single-$\vec{q}$ intravalley pairing where the mean-field superconducting Hamiltonian in the projected basis is given by
\begin{equation}
\begin{split}
\mathcal{H}_{\rm SC}&=\sum_{\vec{k},\sigma}E_{\vec{k}}  f^{\dagger}_{\vec{k},\sigma} f^\pdagger_{\vec{k},\sigma}\\
&\quad+\sum_{\vec{k}}\left[\Delta_{\vec{k}}  f^{\dagger}_{\vec{k}+\vec{q}/2,\uparrow} f^{\dagger}_{-\vec{k}+\vec{q}/2,\downarrow}+\text{H.c.}\right],
\end{split}
\label{eq:HSC_projected}
\end{equation}
where $\Delta_{\vec{k}}$ is the complex-valued superconducting order parameter.
In case the bands that define the normal state for pairing are also spin-polarized and non-degenerate, as is believed to be realized at least in part of the phase diagram of RTG \cite{rtg_chiral_sc_experiment}, the spin index has to be dropped from the fermions in \equref{eq:HSC_projected}. Fermi-Dirac statistics then necessitates that $\Delta_{\vec{k}}= - \Delta_{-\vec{k}}$.

Furthermore, in RTG, the pairing form factor can be distinguished into achiral and chiral, depending on whether the resulting superconductor has trivial or non-zero Chern number. For $\vec{q} =0$, 
 the superconducting order parameter can belong to the  irreducible representation (IR) $A, E$ or $E^*$. In our gauge choice, where the fermionic operators transform trivially under $C_{3z}$, i.e. $f_{\vec{k}}^{\dagger} \rightarrow f_{C_{3z}\vec{k}}^{\dagger}$, the 
 achiral pairing with Chern number zero corresponds to the IR $A$ and the chiral pairing with Chern number $\pm 1$ corresponds to the IRs $E$ and $E^*$, respectively \cite{sedov2025probingsuperconductivitytunnelingspectroscopy}; in the presence of time-reversal symmetry, the latter two would be degenerate. We here choose the ``active'' valley such that the Chern number of the pairing state with order parameter transforming under $E$ has the same sign of the Chern number as the normal state. We consequently analyze the QPI for both topologically trivial $A$ pairing and non-trivial $E$ pairing. 
 
 Two comments are in order: For $\vec{q} \neq 0 $, we exploit the adiabatic continuity to the $\vec{q} =0 $ states to
differentiate the pairing states \cite{OurEnergetics}. Second, note that there is also an ``anti-chiral state''. For the valley chosen, its order parameter would transform under IR $E^*$ and the Chern number of the resulting superconductor would be opposite in sign to that of the active valley. To keep the discussion compact and since \refcite{OurEnergetics} found this state not to be favored for the interactions studied, we will not further consider this form of pairing in the following.

 The impurity potential is treated as in the normal state, but projected onto the active band. For an
impurity located on layer $\beta\in\{\text{t},\text{b}\}$, we use the corresponding impurity matrix $M_{\beta}$ defined
in \equref{eq:ImpurityMatrices}. The projected scattering matrix element becomes
\begin{equation}
W^{(\beta)}_{\vec{k},\vec{k}'}\equiv\langle v^{+}_{\vec{k}}|M_{\beta}|v^{+}_{\vec{k}'}\rangle,
\label{eq:Wkk_def}
\end{equation}
and the corresponding projected impurity Hamiltonian reads
\begin{equation}
\mathcal{H}_{\rm imp}=\sum_{\vec{k},\vec{k}',\sigma} W^{(\beta)}_{\vec{k},\vec{k}'}  f^{\dagger}_{\vec{k},\sigma} f^\pdagger_{\vec{k}',\sigma}.
\label{eq:Himp_projected}
\end{equation}
This band projection is the point where quantum geometry enters the impurity problem most directly through
the momentum dependence of the Bloch spinors $|v^{+}_{\vec{k}}\rangle$. Explicitly, $W^{(\text{t})}_{\vec{k},\vec{k}'} = V_0 A_{\vec{k}} A_{\vec{k}'}$ and $W^{(\text{b})}_{\vec{k},\vec{k}'} = V_0 B^*_{\vec{k}} B_{\vec{k}'}$, emphasizing that $W^{(\text{t})}_{\vec{k},\vec{k}'}$ is real and isotropic whereas $W^{(\text{b})}_{\vec{k},\vec{k}'}$ carries an explicit angular phase $e^{4i(\varphi_{\vec{k}}-\varphi_{\vec{k}'})}$. 
We point out that the disorder matrix elements separate into products of functions, each depending only on one of the indices (i.e., $W_{\vec{k},\vec{k}'}$ has rank 1). This is simply a consequence of the simple form of impurities with only weight on one layer/sublattice in \equref{eq:ImpurityMatrices}. Second, we reiterate that we start from microscopically non-magnetic impurities, i.e., with an associated bare perturbation to the Hamiltonian that preserves time-reversal symmetry. The non-trivial phase factors here instead come from the projection onto the active band and are related to the broken time-reversal symmetry of the Bloch states. The impact of this effect for the stability of the superconductor to a finite density of such defects was discussed in \refcite{2025arXiv251019943S}. We here focus on its consequences for QPI.

Since we are dealing with superconductivity,
 we move to a Nambu basis $\Psi_{\vec{k}}=(f_{\vec{k},\uparrow},  f^{\dagger}_{-\vec{k}+\vec{q},\downarrow})^T$,  to formulate the response. Expressing the clean retarded Green's function $\mathcal{G}^{\text{SC}}_0(\vec{k},\omega)$ in the Nambu basis, associated with \equref{eq:HSC_projected}, the leading order impurity-induced correction to the Nambu Green's function (see \appref{app:QpiSC} for more details) is given by
\begin{equation}
\delta G^\text{SC}(\omega)=[\mathcal{G}^\text{SC}_0(\omega) \mathcal{W} \mathcal{G}^\text{SC}_0(\omega)]_{11},
\label{eq:deltaG_SC}
\end{equation}
where $\mathcal{W}$ contains the impurity matrix elements in
Nambu space, and we have chosen the upper-left (particle-sector) component that enters the LDOS
on a given layer $\alpha\in\{\text{t},\text{b}\}$. Carrying out the Nambu algebra and performing a Fourier transform yields an explicit expression for $\delta G^\text{SC}_{\alpha\beta}(\omega,\vec{r})$
given by
\begin{equation}
\begin{split}
\delta G^{\text{SC}}_{\alpha\beta}(\vec{r}, \omega) &=
\int d^2\vec{k}\int d^2\vec{k}'e^{i(\vec{k}-\vec{k}')\cdot\vec{r}}\\
&\quad\times \langle \alpha|v^{+}_{\vec{k}}\rangle
\frac{\mathcal{N}^{\beta}_{\vec{k},\vec{k}'}(\omega)}{\mathcal{D}_{\vec{k}}(\omega) \mathcal{D}_{\vec{k}'}(\omega)}
\langle v^{+}_{\vec{k}'}|\alpha\rangle,
\end{split}
\label{eq:GFchangeSC}
\end{equation}
where $\alpha\in\{\text{t},\text{b}\}$ labels the measurement layer (STM tip) and $\beta\in\{\text{t},\text{b}\}$ labels the impurity
layer through the choice of $M_{\beta}$ (equivalently, $W^{(\beta)}_{\vec{k},\vec{k}'}$)
with
\begin{align}
\mathcal{D}_{\vec{k}}(\omega) &\equiv \bigl(\omega+i\eta-E_{\vec{k}}\bigr)\bigl(\omega+i\eta+E_{-\vec{k}+\vec{q}}\bigr)\nonumber\\
&\quad-\bigl|\Delta_{\vec{k}-\vec{q}/2}\bigr|^2, \\
\mathcal{N}^{\beta}_{\vec{k},\vec{k}'}(\omega) &\equiv \bigl(\omega+i\eta+E_{-\vec{k}+\vec{q}}\bigr)
W^{(\beta)}_{\vec{k},\vec{k}'}\bigl(\omega+i\eta+E_{-\vec{k}'+\vec{q}}\bigr)\nonumber\\
&\quad- \bigl(W^{(\beta)}_{-\vec{k}+\vec{q},-\vec{k}'+\vec{q}}\bigr)^{*} \Delta_{\vec{k}-\vec{q}/2} \Delta^{*}_{\vec{k}'-\vec{q}/2}. \label{DefinitionOfN}
\end{align}
Clearly, any distinction between the configurations arises entirely from the numerator $ \langle \alpha|v^{+}_{\vec{k}}\rangle\mathcal{N}^{\beta}_{\vec{k},\vec{k}'}(\omega) \langle v^{+}_{\vec{k}'}|\alpha\rangle$. 
This expression satisfies two important consistency checks. First, setting $\Delta_{\vec{k}}=0$ eliminates the $\vec{q}$ dependence and recovers the normal-state response in \equref{eq:GFchangeNS} with $p=p'=+$, as expected. Second, the formalism is gauge invariant: under the Bloch-state gauge transformation $|v^{+}_{\vec{k}}\rangle\to e^{-i\theta_{\vec{k}}}|v^{+}_{\vec{k}}\rangle$, provided the impurity matrix element and order parameter transform as
$W^{(\beta)}_{\vec{k},\vec{k}'}\to e^{i(\theta_{\vec{k}}-\theta_{\vec{k}'})}W^{(\beta)}_{\vec{k},\vec{k}'}$ and
$\Delta_{\vec{k}}\to e^{i(\theta_{\vec{k}+\vec{q}/2}+\theta_{-\vec{k}+\vec{q}/2})} \Delta_{\vec{k}}$, respectively. 

Finally, we define the impurity-induced change in the spectral function analogously to the normal state case as
\begin{equation}
    \delta A^{\text{SC}}_{\alpha\beta}(\vec{r}, \omega)= -\frac{1}{\pi} \text{Im}\left[\delta G^\text{SC}_{\alpha\beta}(\vec{r}, \omega)\right]
    \label{eq:SpectralFunctionSuperconductor}
\end{equation}
Here, the label $\text{SC}\in\{\text{ach},\text{ch}\}$ distinguishes the two different superconducting states, where $\text{ach}$ refers to the achiral state (with IR $A$) and $\text{ch}$ to the chiral one (with IR $E$).

In the following subsections, we apply this formalism to different pairing states for zero- and finite-momentum superconductivity and keep our discussions restricted to the same-layer (tt/bb) configurations for the rest of the manuscript.

\begin{figure}[tb]
\centering
  \includegraphics[width=1.0\linewidth]{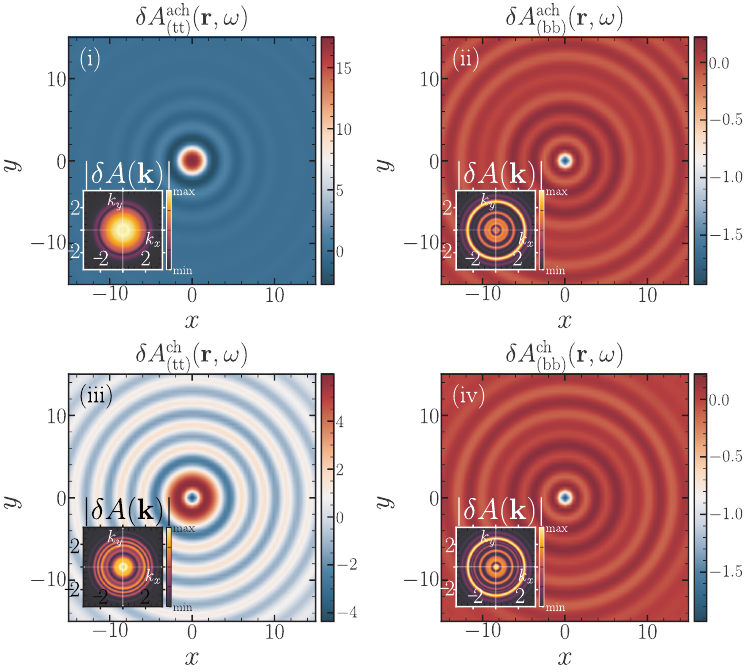}
  \caption{\textbf{Zero momentum superconductivity}. Impurity-induced change in the superconducting spectral function, $\delta A^{\text{SC}}_{\alpha\beta}(\vec{r}, \omega)$ [\equref{eq:SpectralFunctionSuperconductor}] for $\vec{q}=0$ plotted in real space for achiral (first row) and chiral (second row) pairing states in the (i, iii) top-top and (ii, iv) bottom-bottom configurations for $\omega = 1.05, \mu =2$. Insets show the corresponding momentum-space QPI patterns obtained by Fourier transform. All other parameters are the same as \figref{fig:NormalStateQPI}.}
  \label{fig:ZeroMomentumSC}
\end{figure}

\subsection{Zero momentum superconductivity}
We begin by setting $\vec{q}=0$ in \equref{eq:HSC_projected}, corresponding to Cooper pairing with zero center-of-mass momentum. In this case, the BdG quasiparticle bands read $\mathcal{E}_{\vec{k}} = \pm\sqrt{E_{\vec{k}}^2 + |\Delta_{\vec{k}}|^2}$. Since the normal-state dispersion $E_{\vec{k}} = \sqrt{u_0^2 + w_0^2 k^8} - \mu$ satisfies $E_{-\vec{k}} = E_{\vec{k}}$, the factor $\mathcal{D}$ in the denominator of \equref{eq:GFchangeSC} simplifies to
\begin{equation}
\mathcal{D}_{\vec{k}}(\omega) = \bigl(\omega+i\eta\bigr)^2 - \mathcal{E}_{\vec{k}}^2.
\end{equation}
The factor $\mathcal{N}^{\beta}_{\vec{k},\vec{k}'}$ in the numerator at $\vec{q}=0$ takes the form
\begin{equation}
    \mathcal{N}^{\beta}_{\vec{k},\vec{k}'}(\omega)
   = \mathcal{F}_{\vec{k},\omega} W^{(\beta)}_{\vec{k},\vec{k}'}\mathcal{F}_{\vec{k}',\omega} 
    - \bigl(W^{(\beta)}_{-\vec{k},-\vec{k}'}\bigr)^{*} \Delta_{\vec{k}}^{\phantom{*}}\Delta^{*}_{\vec{k}'},
\label{eq:Numer_zeroq}
\end{equation}
where $\mathcal{F}_{\vec{k},\omega} = \bigl(\omega + i\eta + E_{\vec{k}}\bigr)$. We can interpret the first term in the expression to be the normal (particle--particle) contribution where the incoming electron with momentum $\vec{k}'$ scatters off the impurity potential $W^{(\beta)}_{\vec{k},\vec{k}'}$ to momentum $\vec{k}$. The second term is the anomalous (particle--hole) contribution where the incoming electron first converts to a hole through the condensate $\Delta_{\vec{k}'}^{*}$; the hole then scatters off the impurity through $\bigl(W^{(\beta)}_{-\vec{k},-\vec{k}'}\bigr)^{*} $ and finally again converts back to an electron through the condensate $\Delta_{\vec{k}}$. The angular structure of $\mathcal{N}^{b}_{\vec{k},\vec{k}'}(\omega)$ is governed by the interplay between the Bloch-state form factors $A_{\vec{k}}, B_{\vec{k}}$ and the phase of the order parameter. To make this structure explicit, we characterize each contribution (normal and anomalous) by its angular ``winding'' number $l$, defined via the phase factor $e^{il(\varphi_{\vec{k}}-\varphi_{\vec{k}'})}$ it carries.

Before turning to specific pairing states, we establish a general symmetry of the QPI pattern and the key factors at play. The Fourier transform of \equref{eq:GFchangeSC} with respect to $\vec{r}$ and evaluated at $\delta \vec{k}$ yields an integral over $\vec{k}$ at fixed scattering wavevector $\delta\vec{k}$,
\begin{equation}
\delta G^{\text{SC}}_{\alpha\beta}(\omega,\delta\vec{ k}) \propto \int d^2\vec{k}\;\langle\alpha|v^+_{\vec{k}}\rangle 
\frac{\mathcal{N}^{\beta}_{\vec{k},\vec{k}-\delta \vec{k}}(\omega)}{\mathcal{D}_{\vec{k}}(\omega) \mathcal{D}_{\vec{k}-\delta\vec{k}}(\omega)} \langle v^+_{\vec{k}-\delta \vec{k}}|\alpha\rangle.
\label{eq:QPI_kspace}
\end{equation}
We note that, irrespective of the layer configuration and pairing state, the integral depends only on $|\delta\vec{k}|$, i.e., the QPI is rotationally symmetric.  
Another observation is that for same-layer configurations ($\alpha = \beta$), the normal-state contribution to the full integrand is always real and has a total normal winding of $l=0$ for both $\alpha=\text{t}$ and $\alpha=\text{b}$. All differences between configurations and pairing symmetries are therefore encoded in the anomalous term; this is consistent with the conclusion we reached for the normal state, where tt and bb yielded a qualitatively identical profile around the impurity.

\textit{Achiral pairing.---}We first consider the tt configuration with pairing in the IR $A$, and choose a superconducting order parameter given by $\Delta_{\vec{k}} = |\Delta|$. Since $A_{\vec{k}}$ depends only on $|\vec{k}|$ and is real, one has $A_{-\vec{k}} = A_{\vec{k}}$, so the normal and anomalous impurity weights are identical: $(W^{(\text{t})}_{-\vec{k},-\vec{k} + \delta\vec{k}})^* = V_0 A_{\vec{k}} A_{\vec{k} - \delta\vec{k}} = W^{(\text{t})}_{\vec{k},\vec{k} - \delta\vec{k}}$. Both contributions therefore share the same real, isotropic prefactor ($l=0$), and \equref{eq:Numer_zeroq} factorizes as
\begin{equation}
\mathcal{N}^{\text{t}}_{\vec{k},\vec{k} - \delta\vec{k}}(\omega)
= V_0 A_{\vec{k}}A_{\vec{k} - \delta\vec{k}}
\Bigl[\mathcal{F}_{\vec{k},\omega} \mathcal{F}_{\vec{k}-\delta\vec{k},\omega} 
- |\Delta|^2\Bigr]. \label{FactorizationttCase}
\end{equation}
The form factors $A_{\vec{k}}$ form an overall envelope that modulates both contributions equally, as is evident in the inset of panel (i) of \figref{fig:ZeroMomentumSC}, exhibiting a smooth, broad ring as the QPI pattern. For the results in \figref{fig:ZeroMomentumSC} and also all subsequent QPI plots, we first compute the spatial variations around the impurity in a finite square-shaped window, to simulate more closely the situation in experiment. We then perform a discrete Fourier transform after zero padding.

In contrast, for the bb configuration, the situation is completely different. Using $B_{\vec{k}} = N_{\vec{k}} w_0 k^4 e^{-4i\varphi_{\vec{k}}}$ and noting that $B_{-\vec{k}} = B_{\vec{k}}$ the anomalous matrix element evaluates to $(W^{(\text{b})}_{-\vec{k},-\vec{k} +\delta\vec{k}})^* = V_0 B_{\vec{k}} B^*_{\vec{k} - \delta\vec{k}}$, which is the complex conjugate of the normal matrix element. $\mathcal{N}^{\text{b}}_{\vec{k},\vec{k} - \delta\vec{k}}(\omega)$  is then given by  
\begin{equation}
    \mathcal{N}^{\text{b}}_{\vec{k},\vec{k} - \delta\vec{k}}(\omega)
= V_0
\Bigl[B_{\vec{k}}^*B_{\vec{k} - \delta\vec{k}}\mathcal{F}_{\vec{k},\omega} \mathcal{F}_{\vec{k}-\delta\vec{k},\omega} 
 - B_{\vec{k}}B^*_{\vec{k} - \delta\vec{k}}|\Delta|^2\Bigr], \label{NbExpr}
\end{equation}
where the normal and anomalous terms carry opposite phase windings $l=4$ and $l=-4$, respectively. Taking into account the form factors from the measurement layers, we get total normal and anomalous phase windings of $l = 0$ and $l = -8$, respectively, thereby contributing differently to $\delta \mathcal{G}^{\text{SC}}_{\alpha\beta}(\omega,\delta\vec{k})$. Their QPI radial profile is qualitatively distinct from the tt case, as is evident from panel (ii) of \figref{fig:ZeroMomentumSC}, showing suppression instead of enhancement of the LDOS at $\vec{r}=0$. We can further see that the Fourier transform (inset) now displays prominent ring-shaped features. Note that the overall smaller amplitude of the impurity-induced changes in the spectral function for the bb configuration is natural given the interference of the two terms in \equref{NbExpr}. This demonstrates that the sublattice-dependent Bloch form factors alone are sufficient to produce qualitatively different LDOS and QPI signatures on top and bottom layers for an achiral gap, even though $\mathcal{D}_{\vec{k}}$ is identical for both configurations.

\begin{figure*}[tb]
\centering
  \includegraphics[width=1.0\linewidth]{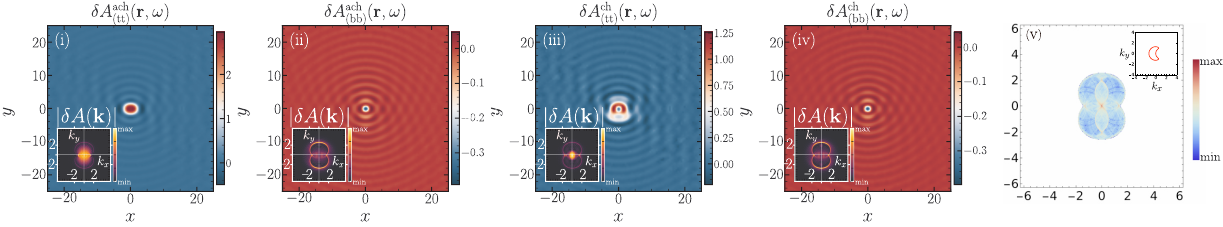}
  \caption{\textbf{Finite momentum superconductivity}. Impurity-induced change in $\delta A^{\text{SC}}_{\alpha\beta}(\vec{r}, \omega)$ in real space and their corresponding QPI patterns (insets) for achiral SC in (i) top-top and (ii) bottom-bottom combination; chiral SC in (iii) top-top and (iv) bottom-bottom combination at $\vec{q}\neq0$. Panel (v) shows the momentum-space density map of the allowed scattering processes, determined by the Fermi surface geometry (inset), which sets the accessible scattering wavevectors visible in the QPI patterns of panels (i)–(iv).}
  \label{fig:FiniteMomentumSC}
\end{figure*}

\textit{Chiral pairing.---}We now turn to the chiral pairing state where we consider the IR $E$, for which we choose $\Delta_{\vec{k}} = |\Delta|e^{-i\varphi_{\vec{k}}}$ as superconducting order parameter. This gap function is odd in $\vec{k}$, $\Delta_{-\vec{k}} = -\Delta_{\vec{k}}$, and, thus, constitutes spin-triplet pairing. In contrast to the achiral case, 
the anomalous contribution in \equref{eq:Numer_zeroq} now involves a product that carries an intrinsic phase winding $e^{-i(\varphi_{\vec{k}}-\varphi_{\vec{k}'})}$ with $l=-1$.

For the tt configuration, the normal contribution remains real and isotropic, whereas the anomalous term now features the $\vec{k}$-dependent phase factor $e^{-i(\varphi_{\vec{k}}-\varphi_{\vec{k}'})}$ from the superconducting order parameter itself. The two contributions, therefore, no longer share a common prefactor, and the numerator cannot be written as a single form-factor envelope as in the achiral case (for tt), see \equref{FactorizationttCase}. This single winding with $l=-1$ in the anomalous contribution leads to the LDOS for the top layer being radially different from the achiral result, as shown in panel (iii) of \figref{fig:ZeroMomentumSC}. Also the Fourier-transformed QPI signals displays rather different behavior. An immediate consequence of this is that the QPI signal can serve as a direct fingerprint to detect a chiral pairing state and to clearly distinguish it from the achiral one in the tt combination. 

For the bb configuration, both the form-factor windings from $B_{\vec{k}}$ and the order-parameter phase contribute simultaneously to give a net winding of $l=-5$ in the anomalous term $\mathcal{N}_{\vec{k},\vec{k}'}$ which further modifies this asymmetry compared to the achiral bb case, producing a distinct radial profile in the QPI pattern visible in panel (iv) of \figref{fig:ZeroMomentumSC}. Since the contrast to \figref{fig:ZeroMomentumSC}(ii) is more subtle, we conclude that, overall, the most clear difference between chiral and achiral pairing can be observed in the tt configuration. In \tableref{SummaryTable} we list a summary of all the windings in $\mathcal{N}^\beta_{\vec{k},\vec{k}'}$ for the different scenarios.

\begin{table}[tb]
\centering
\begin{tabular}{cccc}
\hline
 IR & Config. & Normal winding $l$ & Anomalous winding $l$\\
\hline
$A$ &  tt & $0$ & $0$ \\
$A$ & bb & $4$ & $-4$  \\
$E$ ($E^*$) & tt  & $0$ & $-1$ ($1$)  \\
$E$ ($E^*$) & bb  & $4$ & $-5$ ($-3$) \\
\hline
\end{tabular}
\caption{Summary of the angular winding numbers $l$ of the normal (normal winding) and anomalous contributions (anomalous winding) to $\mathcal{N}^\beta_{\vec{k},\vec{k}'}$ for the IRs $A$, $E$, and $E^*$ of the superconductor at $\vec{q}=0$. The achiral tt configuration  is the only case that yields identical form factors for normal and anomalous terms.} 
\label{SummaryTable}
\end{table}

\subsection{Finite momentum superconductivity}
Finally, we also consider finite-momentum superconductivity, i.e., $\vec{q} \neq 0$ in \equref{eq:HSC_projected}, that can be favored over $\vec{q} = 0$
since time-reversal symmetry is broken in the normal state. While simultaneous condensation of all three, $C_{3z}$-related, momenta is also possible \cite{sedov2025probingsuperconductivitytunnelingspectroscopy,OurEnergetics,k8sb-rqxf,gil2025chargepairdensitywaves,banerjee2025correlationssuperconductingresistiveanisotropies}, such a state breaks translational symmetry (in gauge-invariant quantities) and can therefore be directly identified from the LDOS of the impurity-free system, as detailed in \refcite{sedov2025probingsuperconductivitytunnelingspectroscopy}. This is why we here focus on single-$\vec{q}$ pairing where this is not the case.
Without loss of generality, we choose $\vec{q} = k^*(1,0)^T$, and present the results in \figref{fig:FiniteMomentumSC} for all the possible combinations of impurity position and pairing.

As anticipated, turning on $\vec{q}$ results in the breaking of continuous rotational symmetry of the continuum model, while preserving translational symmetry. As a result, all configurations now exhibit a directional preference in their QPI patterns.  In the inset of \figref{fig:FiniteMomentumSC}(i), this behavior is clearly seen in the pattern, which shows an elongation in the $\delta k_y$ direction on account of our choice of $\vec{q}$. Another important observation is that the momentum-space QPI maps all display a similar overall structure: a dumbbell-shaped envelope, with the distribution of bright and dark regions within this envelope differing from channel to channel. This feature can be understood from the constant energy contour of BdG quasiparticles \cite{volovikpaper,bogoliubov_FS} at the probing energy, shown in the inset of \figref{fig:FiniteMomentumSC}(v). All allowed finite-momentum transfer vectors arise from scattering processes connecting points on that contour. As illustrated in the main panel of \figref{fig:FiniteMomentumSC}(v), the set of all such momentum transfers forms a dumbbell-shaped region in momentum space. Therefore, the constant-energy contour defines the geometric boundary of the allowed momentum-transfer vectors that can contribute to the QPI signal, explaining why the QPI patterns across different channels share the same overall shape in their envelope. However, the intensity distribution within this allowed region is not determined by the BdG spectrum alone. For a given momentum transfer vector that is energetically allowed, contributions may arise from multiple pairs of $\vec{k}$-points on the constant-energy contour. Whether these contributions will add up or cancel depends on both the structure of the Bloch wave functions and the form of the scattering matrix elements. Thus, while the boundary shape of the QPI patterns are set by the constant-energy contour at the probing energy, the observed channel-dependent variations in the intensity distribution arise from the nontrivial quantum geometry of the Bloch wave functions in our model.

Beyond this universal anisotropy, the qualitative distinctions established for $\vec{q}=0$ persist. 
For the bb configuration, panel (ii) of \figref{fig:FiniteMomentumSC} shows distinct behavior in the pattern compared to the tt configuration, demonstrating that the sublattice-dependent Bloch form factors continue to lead to different features for the two configurations even at finite momentum transfer. This distinction also persists for the chiral order parameter as seen in panels (iii, iv) of \figref{fig:FiniteMomentumSC}.  Importantly, the QPI patterns also continue to enable a clear
distinction between chiral and achiral pairing for the tt configuration, proving the robustness of this signature beyond the zero-momentum limit. 

\begin{figure}[tb]
\includegraphics[width=1.0\columnwidth]{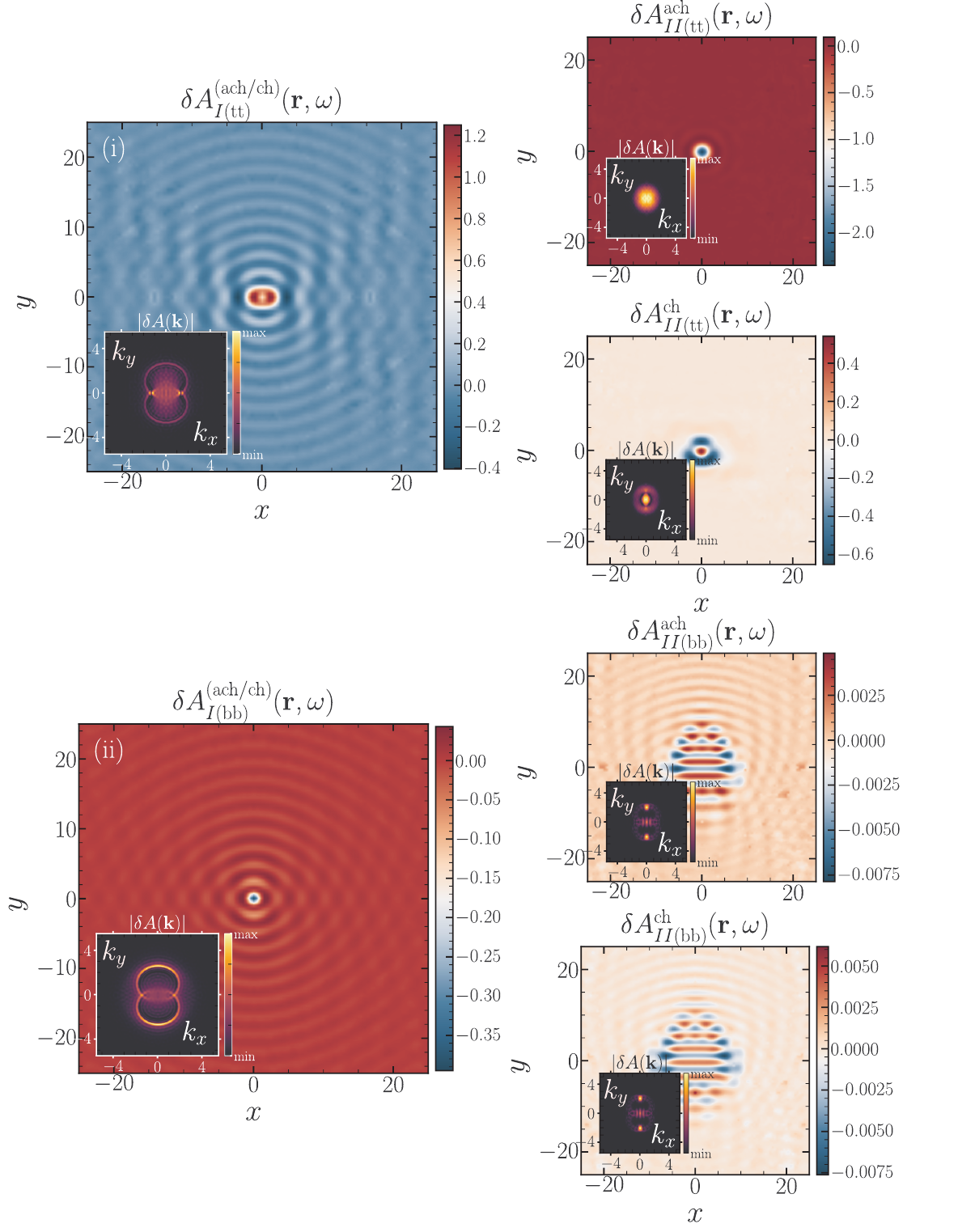}
  \caption{\textbf{Separating normal and anomalous contributions.} Normal and anomalous contributions to $\delta A(\vec{r},\omega)$, denoted as $I$ and $II$, respectively, and corresponding to the first and second term in \equref{DefinitionOfN}. Panel (i) depicts the contributions for the tt configuration corresponding to both achiral and chiral types of pairing. Panel (ii) depicts the same, but for the bb configuration. The corresponding QPI patterns are shown in the insets. While both contributions are of the same order in the tt case (allowing for a distinction between the two pairings), the anomalous one is much smaller than the normal one in the bb configuration.}
  \label{fig:ContributionsNormalAnomalous}
\end{figure}

We confirm this numerically in \figref{fig:ContributionsNormalAnomalous} where we isolate the contributions from the normal and the anomalous parts to $\delta A(\vec{r},\omega)$. In \figref{fig:ContributionsNormalAnomalous}(i) for the tt configuration, the distinguishable (between $E$ and $A$ pairings) contributions mainly arise in the anomalous part. For the bb configuration [\figref{fig:ContributionsNormalAnomalous} (ii)], the normal contributions are $\sim 100$ times larger than the anomalous contributions, and therefore the subtle distinctions in the anomalous parts do not enable a sizeable overall distinction between the two kinds of pairing. 

\section{Conclusion}\label{ConclusionSec}
We have shown that QPI measurements provide a powerful and sensitive probe of the pairing symmetry in superconductors with chiral Bloch bands and that their quantum geometry plays a central role in determining the QPI response. Using a minimal two-band model of rhombohedral graphene as a platform, we have systematically analyzed the impurity-induced change in the spectral function for the tip and impurities being located at different sublattices or layers and for different superconducting order parameters, including both zero- and finite-momentum Cooper pairing.
In the normal state, the non-trivial winding of the Bloch eigenstates leads to a quantum-geometry-induced destructive interference at the impurity site in the mixed-layer (top-bottom/bottom-top) configurations, whereas the same-layer (top-top and bottom-bottom) configurations produce qualitatively similar Friedel oscillations that differ only in their overall magnitude. 

For the zero-momentum superconducting state, we have shown that different pairing states can be clearly distinguished based on their QPI response. In particular, the top-top configuration provides the sharpest and most direct fingerprint to separate between topologically trivial achiral (irreducible representation $A$) and non-trivial chiral (IR $E$) pairing, where the normal scattering contribution is always angularly isotropic ($l=0$), while the anomalous contribution inherits the angular winding directly from the superconducting order parameter. This mismatch in winding numbers produces qualitatively distinct radial profiles in the QPI pattern for chiral versus achiral pairing, making QPI measurements in top-top configuration an ideal experimental tool to distinguish
between pairing symmetries.

We have further shown that for finite-momentum superconductivity, the broken continuous rotational symmetry imprints a universal directional anisotropy on all QPI patterns, whose overall geometric envelope is set by the constant-energy contour of the Bogoliubov quasiparticles. The qualitative distinctions established for zero momentum pairing persist and the top-top configuration continues to exhibit a clear and resolvable difference between chiral and achiral pairing, while the intensity distribution within the allowed momentum-transfer region reflects the non-trivial quantum geometry of the underlying Bloch states.

In summary, our findings demonstrate that QPI is a powerful tool to distinguish different superconducting pairing states in chiral Bloch bands, even though the normal state exhibits low symmetries, including broken time-reversal symmetry. They also reveal how the interplay of the quantum geometry of the Bloch states with the momentum-dependence of the superconducting order parameter can play a central role in  the QPI response.

\begin{acknowledgments}
M.S.S. thanks D.~Sedov for discussions and a previous collaboration on pairing in rhombohedral graphene. P.P.O. acknowledges valuable discussions with R.~M.~Fernandes.
S.B., S.M., and M.S.S.~further acknowledge funding by the European Union (ERC-2021-STG, Project 101040651---SuperCorr). Views and opinions expressed are however those of the authors only and do not necessarily reflect those of the European Union or the European Research Council Executive Agency. Neither the European Union nor the granting authority can be held responsible for them.
\end{acknowledgments}
\newpage
\bibliography{draft_Refs}

\onecolumngrid

\begin{appendix}

\section{QPI in normal state with trigonally warped dispersion}
\label{app:trigonalwarping}

\begin{figure}[h]
\centering
  \includegraphics[width=1.0\linewidth]{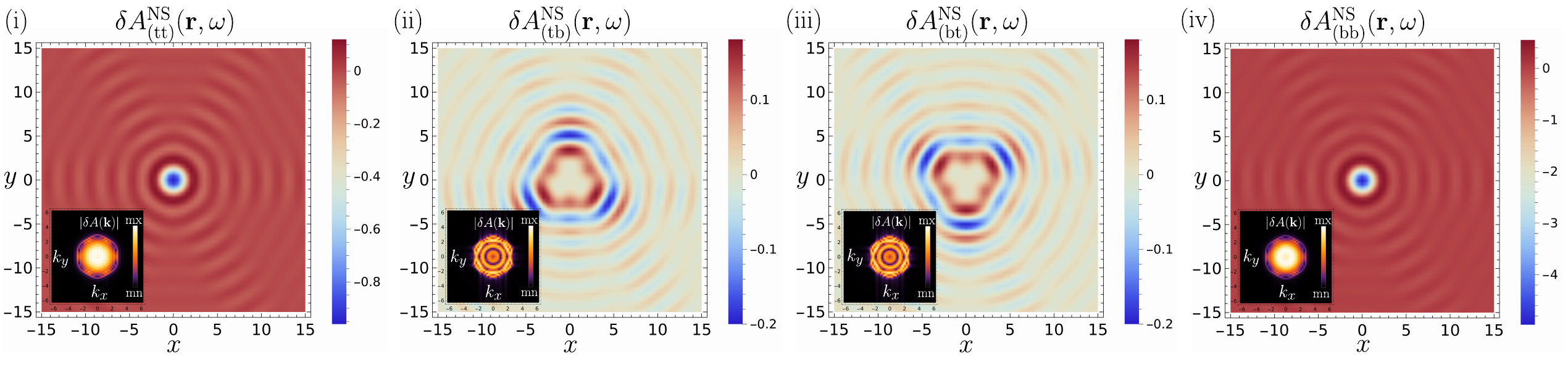}
  \caption{\textbf{Normal state with trigonal warping}. This figure shows the leading-order change in the spectral function $\delta A^{\text{NS}}_{\alpha\beta}(\omega)$ as a function of position, with corresponding QPI maps in momentum space shown in the insets for (i) top--top (tt), (ii) top--bottom (tb), (iii) bottom--top  and (iv) bottom--bottom (bb) configurations. Here we have used $u_{0}=w_{0}=1$, $\mu=0$, $V_0 = 1$, $v_0=0.7$, $w_1=0.3$, $w_2=0.2$, $w_3=0.1$, $\omega = -2$ and $\eta = 0.05$.}
  \label{fig:NSQPIwithtrigwarping}
\end{figure}

In this section, we examine how trigonal warping modifies the QPI pattern in the normal phase. We begin with the effective low-energy Hamiltonian obtained following the prescription of Ref. \cite{koshino2009trigonal}. However, in our current analysis, we employ this Hamiltonian at a phenomenological level. In the $(\psi_{A_1},\psi_{B_4})^T$ basis, we consider that this Hamiltonian takes the form 
\begin{equation}
h(\vec{k}) = 
\begin{pmatrix}
u_0+v_0k^2-\mu && w_0 k^4 + w_1 k|k|^2  +w_2 {k^{\ast}}^2 + w_3 k  \\

w_0 {k^{\ast}}^4 + w_1 k^{\ast}|k|^2 + w_2k^2 +w_3 k^{\ast}&& -u_0+v_0k^2-\mu
\end{pmatrix},
\end{equation}
where $k=k_x +i k_y$. The off-diagonal element now captures the trigonal warping structure. Next, we obtain the eigenstates of this Hamiltonian in a gauge chosen such that it reduces smoothly to those of the rotationally invariant Hamiltonian in the limit of vanishing anisotropy. Finally, using the same procedure outlined in Sec.~\ref{QPINormalState} for calculating the QPI pattern in the normal state, we compute the real-space and momentum-space QPI patterns shown in Fig. \ref{fig:NSQPIwithtrigwarping}.

Fig. \ref{fig:NSQPIwithtrigwarping} clearly shows that the impurity induces oscillations in real space, with corresponding features appearing in momentum space. The inclusion of trigonal-warping terms leads to QPI patterns that display $C_6$ symmetry in the same-layer channels (top-top and bottom-bottom) and $C_3$ symmetry in the mixed-layer channels (top-bottom and bottom-top). While the nature of the QPI signal depends on the layer combination being probed, the main results found in the rotationally invariant case remain intact. In particular, the same-layer channels (tt and bb) display similar QPI patterns characterized by a pronounced peak at the origin, whereas the mixed-layer channels (tb and bt) exhibit a vanishing spectral weight at the origin, as illustrated in Fig. \ref{fig:NSQPIwithtrigwarping}.

\section{QPI in Superconducting State} \label{app:QpiSC}
We work in the band-projected basis with Nambu spinor
\begin{equation}
\Psi_{\vec{k}} = \begin{pmatrix} f_{\vec{k},\uparrow} \\ f^\dagger_{-\vec{k}+\vec{q},\downarrow} \end{pmatrix}.
\end{equation}
The clean BdG Hamiltonian in this basis reads
\begin{equation}
\hat{H}_{\mathrm{BdG}}(\vec{k}) = \begin{pmatrix} E_{\vec{k}} & \Delta_{\vec{k}- \vec{q}/2} \\ \Delta^*_{\vec{k}- \vec{q}/2} & -E_{-\vec{k}+\vec{q}} \end{pmatrix},
\end{equation}
and the impurity Hamiltonian, projected onto the active band for an impurity on layer $\beta$, is given by
\begin{equation}
\mathcal{H}_{\mathrm{imp}} = \sum_{\vec{k},\vec{k}',\sigma} W^{(\beta)}_{\vec{k},\vec{k}'}f^\dagger_{\vec{k},\sigma}f_{\vec{k}',\sigma},
\end{equation}
with $W^{(\beta)}_{\vec{k},\vec{k}'} = \langle v^+_{\vec{k}}|M_\beta|v^+_{\vec{k}'}\rangle$.
The retarded Green's function for the clean superconductor reads
\begin{equation}
\mathcal{G}^{\mathrm{SC}}_0(\vec{k},\omega) = \bigl[(\omega+i\eta) \tau_0 - \hat{H}_{\mathrm{BdG}}(\vec{k})\bigr]^{-1}.
\end{equation}
Inverting the $2\times 2$ matrix explicitly yields
\begin{equation}
\mathcal{G}^{\mathrm{SC}}_0(\vec{k},\omega) = \frac{1}{\mathcal{D}_{\vec{k}}(\omega)}
\begin{pmatrix}
\omega+i\eta+E_{-\vec{k}+\vec{q}} & \Delta_{\vec{k}- \vec{q}/2} \\
\Delta^*_{\vec{k}- \vec{q}/2} & \omega+i\eta-E_{\vec{k}} 
\end{pmatrix},
\end{equation}
where the denominator is given by
\begin{equation}
\mathcal{D}_{\vec{k}}(\omega) = (\omega+i\eta - E_{\vec{k}})(\omega+i\eta+E_{-\vec{k}+\vec{q}}) - |\Delta_{\vec{k}-\vec{q}/2}|^2.
\end{equation}
In the Nambu basis, the impurity scattering vertex for an impurity on layer $\beta$ becomes
\begin{equation}
\hat{V}^{(\beta)}_{\vec{k},\vec{k}'} = \begin{pmatrix} W^{(\beta)}_{\vec{k},\vec{k}'} & 0 \\ 0 & -\bigl(W^{(\beta)}_{-\vec{k}+\vec{q},-\vec{k}'+\vec{q}}\bigr)^* \end{pmatrix}.
\label{eq:NambuImpurity}
\end{equation}

To first order in the impurity potential, the correction to the Nambu Green's function is given by the standard Born approximation as 
\begin{equation}
\delta\mathcal{G}^{\mathrm{SC}}(\vec{k},\vec{k}',\omega) = \mathcal{G}^{\mathrm{SC}}_0(\vec{k},\omega)\hat{V}^{(\beta)}_{\vec{k},\vec{k}'}\mathcal{G}^{\mathrm{SC}}_0(\vec{k}',\omega).
\end{equation}
We are interested in the $(1,1)$ component of $\delta \mathcal{G}^{\mathrm{SC}}$, i.e., the correction to the particle Green's function. Performing the $2\times 2$ matrix multiplication results in 
\begin{align}
\bigl[\delta\mathcal{G}^{\mathrm{SC}}\bigr]_{11} &= \bigl[\mathcal{G}_0\bigr]_{11}W^{(\beta)}_{\vec{k},\vec{k}'}\bigl[\mathcal{G}_0\bigr]_{11}' - \bigl[\mathcal{G}_0\bigr]_{12}\bigl(W^{(\beta)}_{-\vec{k}+\vec{q},-\vec{k}'+\vec{q}}\bigr)^*\bigl[\mathcal{G}_0\bigr]_{21}',
\end{align}
where primes denote evaluation at $\vec{k}'$, and we used $[\mathcal{G}_0]_{12} = \Delta_{\vec{k}}/\mathcal{D}_{\vec{k}}$ and $[\mathcal{G}_0]_{21}' = \Delta^*_{\vec{k}'}/\mathcal{D}_{\vec{k}'}$.
Writing out the components explicitly, we find
\begin{equation}
\bigl[\delta\mathcal{G}^{\mathrm{SC}}\bigr]_{11} = \frac{1}{\mathcal{D}_{\vec{k}}\mathcal{D}_{\vec{k}'}}\biggl[
(\omega+i\eta+E_{-\vec{k}+\vec{q}})W^{(\beta)}_{\vec{k},\vec{k}'} (\omega+i\eta+E_{-\vec{k}'+\vec{q}})
- \bigl(W^{(\beta)}_{-\vec{k}+\vec{q},-\vec{k}'+\vec{q}}\bigr)^*\Delta_{\vec{k}-\vec{q}/2}\Delta^*_{\vec{k}'-\vec{q}/2}
\biggr].
\end{equation}
We finally identify the numerator as stated in Eq.~\eqref{DefinitionOfN} of the main text
\begin{equation}
\mathcal{N}^{\beta}_{\vec{k},\vec{k}'}(\omega) = (\omega+i\eta+E_{-\vec{k}+\vec{q}})W^{(\beta)}_{\vec{k},\vec{k}'}(\omega+i\eta+E_{-\vec{k}'+\vec{q}})
- \bigl(W^{(\beta)}_{-\vec{k}+\vec{q},-\vec{k}'+\vec{q}}\bigr)^*\Delta_{\vec{k}-\vec{q}/2}\Delta^*_{\vec{k}'-\vec{q}/2}\,.
\end{equation}

\section{Projection onto measurement layer and Fourier transformation}

The LDOS correction measured on layer $\alpha$ involves projecting from band space back to layer space using the Bloch form factors:
\begin{equation}
\delta G^{\mathrm{SC}}_{\alpha\beta}(\omega,\vec{k},\vec{k}') = \langle\alpha|v^+_{\vec{k}}\rangle\frac{\mathcal{N}^{\beta}_{\vec{k},\vec{k}'}(\omega)}{\mathcal{D}_{\vec{k}}(\omega)\mathcal{D}_{\vec{k}'}(\omega)}\;\langle v^+_{\vec{k}'}|\alpha\rangle.
\end{equation}
Fourier transforming to real space (with the impurity at the origin) then yields
\begin{equation}
\delta G^{\mathrm{SC}}_{\alpha\beta}(\omega,\vec{r}) = \int d^2\vec{k}\int d^2\vec{k}'e^{i(\vec{k}-\vec{k}')\cdot\vec{r}}\langle\alpha|v^+_{\vec{k}}\rangle\frac{\mathcal{N}^{\beta}_{\vec{k},\vec{k}'}(\omega)}{\mathcal{D}_{\vec{k}}(\omega)\mathcal{D}_{\vec{k}'}(\omega)}\langle v^+_{\vec{k}'}|\alpha\rangle.
\label{eq:deltaG_Appendix}
\end{equation}
The phase $e^{i(\vec{k}-\vec{k}')\cdot\vec{r}}$ encodes the spatial oscillations of the Friedel-like modulations around the impurity. The LDOS correction follows as $\delta A^{\text{SC}}_{\alpha\beta}(\vec{r}, \omega) = -\frac{1}{\pi} \mathrm{Im}\delta G^{\mathrm{SC}}_{\alpha\beta}(\omega,\vec{r})$.

\section{Fourier transform of real-space LDOS modulation to QPI signal}
 
Starting from Eq.~\eqref{eq:deltaG_Appendix} and changing variables to $\delta\vec{k} = \vec{k}-\vec{k}'$ (scattering wavevector) as well as integrating over $\vec{r}$, or equivalently Fourier transforming directly, results in 
\begin{equation}
\delta G^{\mathrm{SC}}_{\alpha\beta}(\omega,\delta\vec{k}) \propto \int d^2\vec{k} \langle\alpha|v^+_{\vec{k}}\rangle \frac{\mathcal{N}^{\beta}_{\vec{k},\vec{k}-\delta\vec{k}}(\omega)}{\mathcal{D}_{\vec{k}}(\omega) \mathcal{D}_{\vec{k}-\delta\vec{k}}(\omega)}\langle v^+_{\vec{k}-\delta\vec{k}}|\alpha\rangle.
\end{equation}
This is the QPI signal: the Fourier-space image of the impurity-induced LDOS modulations, which is the quantity directly comparable to STM experiments.

\end{appendix}

\end{document}